\documentclass{article}
\usepackage[utf8]{inputenc}
\usepackage{graphicx}
 \usepackage{url}
 \usepackage{hhline}
\usepackage{caption}
\usepackage{booktabs}
\usepackage{ragged2e}
\usepackage{tabularx}
\usepackage{xcolor}

\newcolumntype{L}{>{\raggedright\arraybackslash}X}
\newcolumntype{C}{>{\centering\arraybackslash}X}
\usepackage{comment}

\title{Unraveling the Butterfly Effects in Social Dynamics: Insights from Agent-Based Modeling}

\author{
  Sabzian, Hossein  \\
  \texttt{ \small hossein.sabzian@maine.edu} \\
  \texttt{ \small University of Maine, ME, USA} \\
  \and
  Shahriari, Nima \\
  \texttt{\small Nima.x.Shahriari@kp.org} \\
  \texttt{ \small Kaiser Permanente, CA, USA} \\
    \and
   G. Nejad, Mohammad \\
  \texttt{\small mnejad@fordham.edu} \\
  \texttt{ \small Fordham University, NY, USA} \\
}

\date{\today}

\begin{document}

\maketitle

\textbf{ Abstract }

The complex interplay between human societies and their environments has long been a subject of fascination for social scientists. Utilizing agent-based modeling, this study delves into the profound implications of seemingly inconsequential variations in social dynamics. Focusing on the nexus between food distribution, residential patterns, and population dynamics, our research provides an analysis of the long-term implications of minute societal changes over the span of 300 years. 

Through a meticulous exploration of two scenarios, we uncover the profound impact of the butterfly effect on the evolution of human societies, revealing the fact that even the slightest perturbations in the distribution of resources can catalyze monumental shifts in residential patterns and population trajectories. This research sheds light on the inherent complexity of social systems and underscores the sensitivity of these systems to subtle changes, emphasizing the unpredictable nature of long-term societal trajectories. The implications of our findings extend far beyond the realm of social science, carrying profound significance for policy-making, sustainable development, and the preservation of societal equilibrium in an ever-changing world.


\section{Introduction}

We inhabit a dynamic world, where each day unfolds with a spectrum of phenomena of varying intricacies. Ranging from the dissemination of viruses within our environment and the circulation of rumors through human societies to the establishment of diverse social norms and the emergence of groundbreaking technologies, these instances all exemplify the manifestations within our world.

We live in a vibrant, ever-evolving world, one replete with a tapestry of multifaceted phenomena that unfold daily. From the intricate diffusion of viral entities in our surroundings to the delicate dissemination of rumors across the societal fabric, and from the intricate web of established social norms to the continual emergence of groundbreaking technological innovations, these instances collectively mirror the complexity inherent in our world.

To unravel the intricacies of such phenomena, social scientists have traditionally adopted a reductionist lens, disassembling and simplifying these multifaceted dynamics into lower-level variables. Leveraging a suite of equations, including partial and ordinary differential equations, they construct models to illustrate the interplay among these variables. Yet, this reductionist perspective, often termed equation-based modeling (EBM), confronts inherent limitations when confronted with the intricacies of real-world complexities. In the context of social evolution, adherents of this approach tend to segment the studied system into isolated factors, often assuming boundless rationality and an abundance of information. Methodologically, this approach often overlooks critical elements such as historical context, agents' adaptive capacities, the evolutionary nature of these components, and the profound influence of environmental network effects.

In response to the shortcomings of the reductionist framework, the concept of complex adaptive systems (CAS) has gained considerable traction over the past decades. In contrast to reductionism, CAS adopts an organic and dynamic lens for understanding socio-economic phenomena, including the trajectory of social evolution. Within this paradigm, agents (individual actors) possess not only bounded rationality but also a potent adaptive and learning proclivity. These agents engage in a dynamic, intricate network of interactions within the social fabric of their immediate environment. According to this perspective, socio-economic phenomena emerge from the complex interplay and decision-making behavior of constituent actors, challenging reduction to isolated components of society. Agent-based modeling (ABM), a primary methodology for elucidating the intricacies of complex adaptive systems (CAS) , has garnered significant attention among contemporary scholars for its ability to capture how the fundamental behavioral rules governing agents and their localized interactions at the micro level give rise to intricate and often unexpected patterns at the macro level, reflecting the collective tapestry of society.

This paper unfolds in several parts, with Part 2 delving into the theoretical underpinnings, encompassing a thorough exploration of chaos theory, its intersection with complexity theory, an analysis of the properties inherent in complex adaptive systems (CAS), and an in-depth discussion on the complex interplay between society and CAS. Part 3 outlines the methodology and various modeling approaches employed in studying CAS, while Part 4 presents a hypothetical social simulation example. Subsequently, Part 5 delves into the simulation results, unraveling the implications derived from the application of ABM in a societal context. Finally, Part 6 synthesizes the paper's key insights, offering a nuanced conclusion that reflects on the intricacies of modeling societal dynamics within the complex adaptive systems framework and the implications of understanding butterfly effects for social scientists.

\section{Theoretical backgrounds}

\subsection{Chaos theory}  

Chaos theory, a multidisciplinary field of scientific exploration and a branch of mathematics, delves into the underlying patterns and deterministic laws of dynamical systems that exhibit a high sensitivity to initial conditions. Initially perceived as systems with entirely random states of disorder and irregularities, chaos theory aims to understand the behavior of these complex systems \cite{wikichaos}. Its foundation lies in the mathematical representation of a range of phenomena within the domain of dynamics, a branch of physics concerned with the impact of forces on object motion. The classical theory of dynamics, epitomized by Newton's work on celestial motions, serves as a fundamental archetype for the field \cite{oestreicher2022history}.

\subsection{Chaos theory and complexity theory connection}  

Chaotic systems and complex systems both fall under the category of nonlinear dynamical systems \cite{rickles2007simple}. A complex system comprises multiple interacting components that form an irreducible whole greater than the sum of its parts \cite{sabzian2018review}. While there are shared characteristics between chaotic and complex systems, the two concepts diverge significantly. Chaos denotes the generation of intricate, aperiodic, seemingly random behavior from the iteration of a simple rule, as defined precisely in mathematical terms. Complexity, on the other hand, signifies the emergence of rich, collective dynamic behavior resulting from simple interactions among numerous subunits. It is important to note that chaotic systems are not necessarily complex, and complex systems are not necessarily chaotic \cite{bertuglia2005nonlinearity}.

The interactions among the subunits of a complex system give rise to properties within the system that cannot be reduced solely to the subunits themselves or deduced directly from their interactions. Such properties are known as emergent properties. This creates an upward or generative hierarchy of levels, wherein each level of organization determines the level above it, thereby shaping the features of the subsequent level \cite{rickles2007simple}.

A vital aspect common to both chaos and complexity, due to their nonlinear nature, is the sensitivity to initial conditions. This implies that even with a minimal difference in initial states and adherence to the same simple rules, two states can follow vastly different trajectories over time. Such sensitivity poses a challenge in predicting the evolution of a system, as it demands the accurate description of the system's initial state. In practice, errors during this process can compound over time, leading to substantial discrepancies. Consequently, replicating initial conditions becomes challenging in various trial and intervention scenarios \cite{philippe2004evidence}.

In chaos theory, this sensitivity to initial conditions is referred to as the butterfly effect, wherein small changes in the state of a deterministic nonlinear system can yield substantial disparities in subsequent states. This core characteristic of chaotic systems underscores the idea that minor alterations or events within a complex system can trigger significant repercussions elsewhere. Coined by mathematician and meteorologist Edward Lorenz, the butterfly effect metaphorically illustrates how the flapping of a butterfly's wings in Brazil could potentially set off a chain of events leading to a tornado in Texas. This concept highlights the intricate dependence on initial conditions in chaotic systems, revealing how seemingly insignificant actions or events can have widespread and unforeseeable consequences in dynamic and complex systems \cite{lorenz1972predictability}.

\subsection{CAS a a type of complex system} 

Complex adaptive systems (CAS) are special cases of complex systems that are adaptive in that they have the capacity to change and learn from experience. Examples of complex adaptive systems include the stock market, social insect and ant colonies, the biosphere and the ecosystem, the brain and the immune system, the cell and the developing embryo, the cities, manufacturing businesses and any human social group-based endeavor in a cultural and social system such as political parties or communities\cite{skrimizea2019complexity}.A CAS has some fundamental characteristics such as \textit{1: multiplicity and heterogeneity of constituent components}, \textit{2: Non-linear interactions}, \textit{3: Learnability and adaptability}, \textit{4: Non-ergodicity}, \textit{5: Self-organization}, \textit{6: Emergence}, \textit{7: Co-evolution} and \textit{8: Far from equilibrium} \cite{sabzian2019theories}. Naturally the non-ergodic property of a CAS indicate a high degree of sensitivity to initial conditions. 

Complex adaptive systems (CAS) have found applications in various fields, showcasing the versatility and significance of this theoretical framework. Some of these applications can be:

\begin{enumerate}

    \item \textit{Ecology and Evolutionary Biology:} Complex adaptive systems have been applied to study ecological systems and evolutionary dynamics. The interplay of multiple factors, such as species interactions, environmental changes, and population dynamics, can be better understood through the lens of complex adaptive systems\cite{levin1998ecosystems}
    
    \item \textit{Economics and Social Sciences:} CAS has been utilized to analyze economic systems, market dynamics, and social networks. These applications provide insights into the emergence of collective behavior, market trends, and the impact of individual decisions on macro-level outcomes\cite{arthur2013complexity}
    \item \textit{Public Health and Epidemiology:} CAS models have been employed to understand the spread of diseases, analyze healthcare systems, and predict epidemic patterns. These applications aid in the formulation of effective public health policies and strategies for disease control and prevention\cite{epstein2009modelling}
    
    \item \textit{Urban Studies and Transportation Planning:} Complex adaptive systems have been used to study traffic flows, urban development patterns, and transportation network dynamics. These applications help in optimizing urban infrastructure, improving transportation systems, and addressing congestion and mobility challenges in cities\cite{batty2007cities}
    
    \item \textit{Computer Science and Artificial Intelligence:} CAS concepts have been integrated into the development of artificial intelligence, multi-agent systems, and self-organizing algorithms.These applications facilitate the design of intelligent and adaptive systems capable of learning, problem-solving, and decision-making in dynamic and unpredictable environments\cite{mitchell2009complexity, sabzian2020modeling}

\end{enumerate}

\subsection{Society as a CAS}

The characterization of society as a complex adaptive system (CAS) is rooted in its intricate dynamics and the emergent behaviors that materialize from the intricate interactions among diverse social agents and institutions. This understanding has garnered extensive traction in the annals of social science and complex systems literature.
A comprehensive examination of complex adaptive systems, underscoring the manifestation of complex and adaptive behaviors across multiple domains, including society, has been eloquently elucidated by Holland \cite{holland1992complex}.Delving into the emergence of properties within complex systems, inclusive of social systems, the intricate interplay among individual agents has been studied, shedding light on the parallels discernible between natural and social systems \cite{johnson2002emergence}. Furthermore, the exploration of computational models of social life highlights the profound alignment between the features of social systems and the properties  of complex adaptive systems \cite{miller2009complex}.By delving into the dynamics of social movements through the lens of complexity theory, works such as that by Chesters et al. showcase how the actions and interactions of a diverse array of social agents contribute to the genesis and evolution of these movements \cite{chesters2006complexity}. Additionally, the exploration of the concept of social emergence provides a nuanced perspective, replete with case studies and theoretical underpinnings that illuminate the nuanced understanding of society as an emergent complex system \cite{sawyer2005social}.

\section{Methodology}

\subsubsection{Modeling approaches of complex adaptive systems}

Approaches for modeling complex adaptive systems include various methodologies and techniques. These approaches can be technically divided into Equation-based modeling (EBM )and Agent-based modeling (ABM). These modeling techniques offer powerful tools for analyzing and simulating the behavior of complex adaptive systems, providing insights into their dynamics, structure, and emergent properties. These techniques help in understanding the dynamics and emergent properties of complex systems\cite{sabzian2018review, miller2009complex}.

\textit{Equation-based modeling(EBM)} techniques involve the use of mathematical equations to describe the behavior of complex systems. These techniques are particularly valuable for capturing the interactions between various components of a system and understanding how changes in one component can affect the entire system. Equation-based models often rely on mathematical representations derived from fundamental principles, physical laws, or empirical observations. Equation-based modeling techniques provide a robust framework for analyzing the behavior of complex systems, offering insights into the underlying dynamics and interactions among various system components. They are essential tools in fields such as physics, engineering, ecology, and economics for understanding the complexities of natural and artificial systems. The most widely used form of Equation-based modeling techniques are differential equations used in System dynamics modeling and graph algorithms used in Network analysis\cite{sterman2002system, barabasi2013network}

\textit{Agent-based modeling (ABM)}is a computational modeling technique that simulates the actions and interactions of autonomous agents to understand how these interactions give rise to complex system behavior. Agents can represent individuals, groups, or other entities, and the model captures how their individual behaviors and decision-making processes lead to emergent phenomena at the system level. It is commonly used to study the behavior of complex systems in various fields, including economics, biology, and social sciences\cite{epstein1996growing}.

\subsubsection{ABM vs EBM in modeling CAS}

Agent-based models (ABMs) are often preferred over equation-based models (EBMs) when modeling social complex systems due to their ability to capture emergent behaviors and non-linear interactions among individual agents within a system. Here are some key reasons why ABMs can be more effective for social complex systems:

\begin{enumerate}
    \item \textit{Individual Heterogeneity:} ABMs allow for the representation of diverse characteristics, behaviors, and decision-making processes among individual agents, thus enabling a more realistic portrayal of social dynamics.
    \item \textit{Non-Linearity and Emergence:} ABMs are adept at capturing the non-linear and emergent properties of social systems, which are often challenging to represent using traditional mathematical equations. They can simulate how complex global behaviors emerge from simple local interactions among agents.
    \item \textit{Complex Interactions and Feedback Loops:} ABMs can incorporate intricate interactions and feedback loops among agents, facilitating the examination of how local interactions lead to the emergence of global patterns and behaviors.
    \item Realism and Flexibility: ABMs provide a more flexible and realistic framework to model social systems as they can incorporate complex real-world scenarios, allowing researchers to study social phenomena in a more natural and dynamic environment.
    \item \textit{Behavioral Insights:} By focusing on individual-level behaviors and decision-making processes, ABMs enable researchers to gain insights into the underlying mechanisms that drive the behavior of social systems, offering a deeper understanding of social phenomena.
    \item \textit{Adaptability and Change:} ABMs are well-suited for modeling adaptive behaviors and changes in social systems over time, making them ideal for studying the dynamic nature of social interactions and the effects of varying environmental conditions.
\end{enumerate}

When dealing with the complexities and dynamics of social systems, the agent-based approach provides a more comprehensive and nuanced perspective, allowing researchers to explore the intricate relationships and behaviors that shape social phenomena.

\section{A hypothetical social simulation example }

Drawing on the definition of a complex adaptive system (CAS), society is a clear example of a CAS. It is a system of multiple agents which are not only able to learn and change their expectations (adaptive expectations), but also generate a new amount of experience and knowledge and transmit it to other society members over time. Therefore, every phenomenon in the context of society is the result of the interactions of its members in a historical context. Regarding the historical context of a society shows that the as-is situation of two societies should never be examined without considering their past and initial conditions. Even if the two societies have historical similarities in many ways, only one slightest institutional or geographical difference in the past can change their future course. Using a hypothetical example, we would elaborate on how an extremely slight difference in initial states of two somewhat identical societies can bring about two remarkably different trajectories.

\subsection{Description of Societies }

\begin{figure}[!ht]
\includegraphics[width=5in]{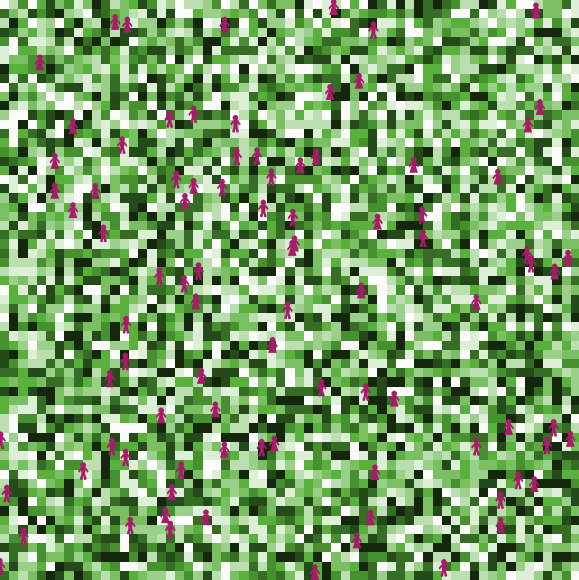}
\caption{Geographical distribution of society A}
\label{fig:SociA}
\end{figure}

\begin{figure}[!ht]
\includegraphics[width=5in]{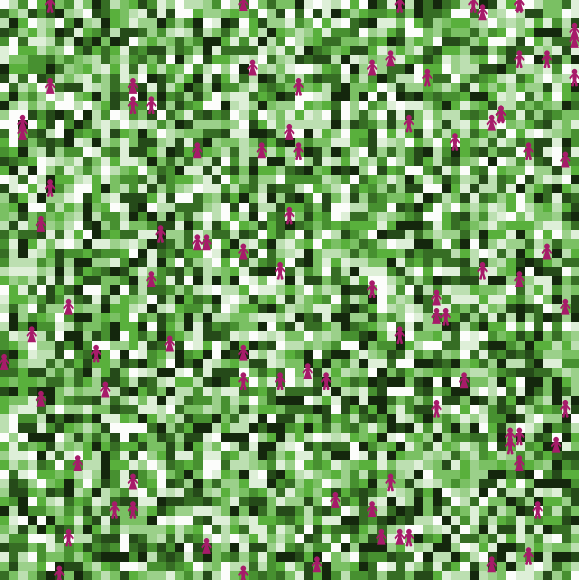}
\caption{Geographical distribution of society B}
\label{fig:sociB}
\end{figure}

We consider two societies, labeled A and B. All initial parameters, as outlined in Table~\ref{tab:initial}, were identical for both societies 300 years ago, with the exception of their geographical distributions (see Figure~\ref{fig:SociA} and Figure~\ref{fig:sociB}).\footnote{The model has been implemented in NetLogo 6.2.0 and is accessible through the following link: \url{https://modelingcommons.org/browse/one_model/7313#model_tabs_browse_info}}

\begin{table}[!h]
\renewcommand{\arraystretch}{1.3}
\caption{The initial conditions of societies in the beginning of the historical period 300 years ago}
\label{tab:initial}
\centering
\begin{tabularx}{\columnwidth}
{ @{\hspace{.2em}} >{\hsize=1.5\hsize}L *{4}{>{\hsize=0.875\hsize}C} @{} }
\toprule
\centering\arraybackslash\textbf{Parameter} & \textbf{Value in Society A} & \textbf{Value in society B}  \\
\midrule
\hspace{.5em} Initial population                                          & 90    & 90    \\
\hspace{.5em} Number of males                                              & 42    & 42    \\
\hspace{.5em} Number of females                                            & 48    & 48      \\
\hspace{.5em} Number of people owning quadruped animals (for riding)       & 0.3    & 0.3    \\
\hspace{.5em} The minimum age of reproduction                              & 18 years    & 18 years    \\
\hspace{.5em} The maximum age of reproduction                               & 33 years    & 33 years     \\
\hspace{.5em} The probability of a successful birth giving (delivery)       & 0.5    & 0.5      \\
\hspace{.5em} Critical threshold of overpopulation (Carrying capacity)     & 1000 persons    & 1000 persons     \\
\hspace{.5em} The amount of energy needed for finding food           & 0.75 unit    & 0.75 unit      \\
\hspace{.5em} The amount of energy needed for finding a partner            & 2 units    & 2 units     \\
\hspace{.5em} The amount of energy gained from eating food                  & 3 units    & 3 units     \\
\hspace{.5em} Probability of plant regrowth                                   & 0.1    & 0.1      \\
\hspace{.5em} The volume of food (plants) available in the Northeast region   & 4145    & 4145     \\
\hspace{.5em} The volume of food (plants) available in the Northwest region   & 4260    & 4260    \\
\hspace{.5em} The volume of food (plants) available in the Southeast region    & 4556    & 4556    \\
\hspace{.5em} The volume of food (plants) available in the Southwest region     & 4432    & 4432     \\
\bottomrule
\end{tabularx}
\end{table}

It is evident that the initial conditions are alike in the two societies. However, their geographical distributions differ, leading to varying patterns of population distribution. We aim to explore the following inquiries:

\begin{enumerate}
\item \textit{What are the expected outcomes of social interactions in each society over a 300-year period?}
\item \textit {At what stage will the population distribution stabilize?}
\end{enumerate}

\section{Simulation results}

\begin{figure}[!ht]
\includegraphics[width=6.5in]{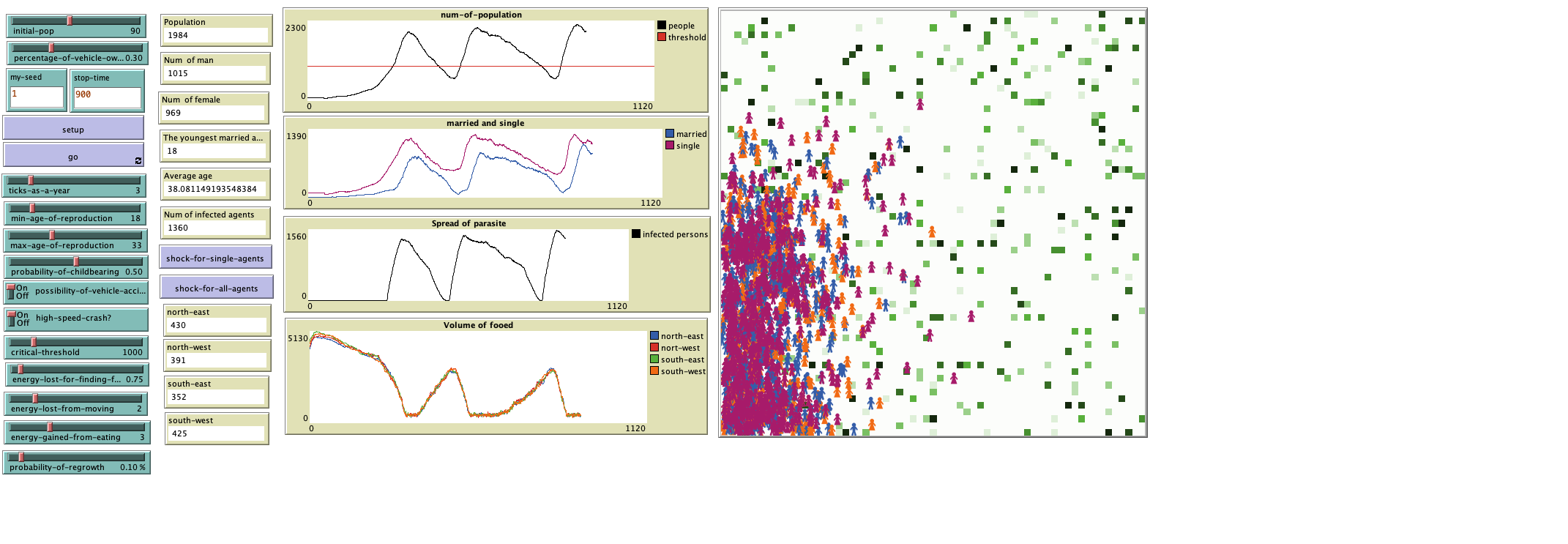}
\caption{Geographical distribution of society A after 300 years}
\label{fig:SociAinterface}
\end{figure}

\begin{figure}[!ht]
\includegraphics[width=5in]{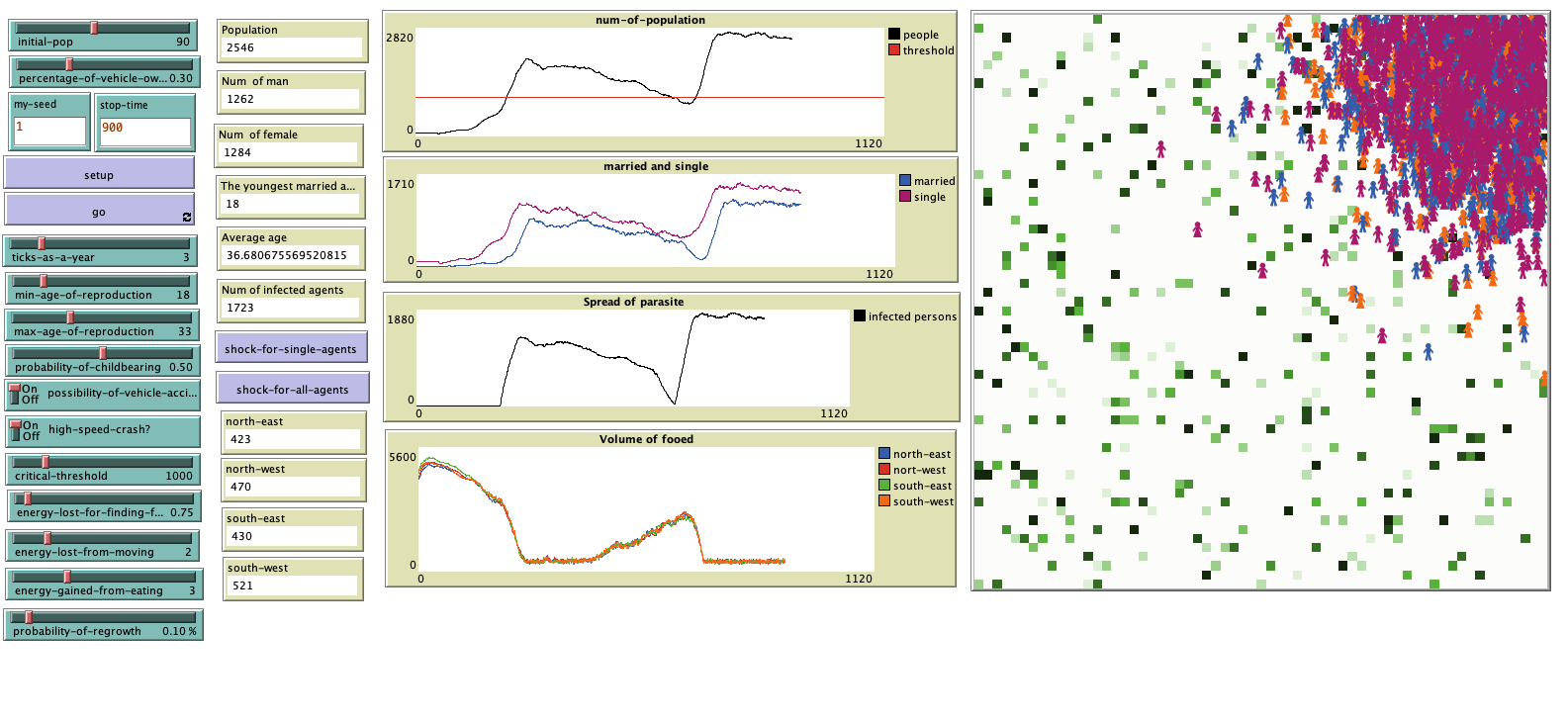}
\caption{Geographical distribution of society B after 300 years}
\label{fig:SociBinterface}
\end{figure}

Simulation results of two societies are shown in Figure~\ref{fig:SociAinterface} and Figure~\ref{fig:SociBinterface}. As it can be seen, the distribution pattern of society A has got concentrated in the southwest part, while the distribution pattern of society B has concentrated in the southeast.

Numerical results of simulation for these two societies in a 300-year period based on demographic and food results are presented in Table ~\ref{tab:change}.

{\small
\begin{table}[!h]
\renewcommand{\arraystretch}{1.1}
\caption{Changes in societies according to demographic and food parameters after 300 years.}
\label{tab:change}
\centering
\begin{tabularx}{\columnwidth}
{ @{\hspace{.2em}} >{\hsize=1.5\hsize}L *{6}{>{\hsize=0.875\hsize}C} @{} }
\toprule
\centering\arraybackslash\textbf{Parameter} & \textbf{Value in Society A} & \textbf{Society A after 300 years} & \textbf{\small Change in Society A} & \textbf{Value in Society B} & \textbf{Society B after 300 years} & \textbf{\small Change in Society B}   \\
\hline
\hspace{.5em} Initial Population & 90 & 1984 & 2104.44\% & 90 & 2546 & 2728.88\%  \\
\hline
\hspace{.5em} Number of males   & 42 & 1015 & 2316.66\% & 42 & 1262 & 2904.76\% \\
\hline
\hspace{.5em} Number of females  & 48  & 969 & 1918.75\% & 48  & 1284 &  2575\%  \\
\hline
\hspace{.5em} Volume of food (plants) in Northeast & 4145 & 430  &  \small-89.62\% & 4145 & 423 &  \small-89.79\%   \\
\hline
\hspace{.5em} Volume of food (plants) in Northwest & 4260 & 391  &  \small-90.82\% & 4260 & 470 & \small-88.96\%  \\
\hline
\hspace{.5em} Volume of food (plants) in Southeast & 4556 & 352 & \small-92.27\% & 4556  & 430 & \small-90.56\%   \\
\hline
\hspace{.5em} Volume of food (plants) in Southwest  & 4432 & 425 & \small-90.41\% & 4432 & 521 &  \small-88.24\%  \\
\bottomrule
\end{tabularx}
\end{table}
}

Table~\ref{tab:change} compares the changes in various parameters between societies A and B after a period of 300 years. As it can be seen, for four first variables, the society has exhibited a higher change while for last three variables, the society A has show a higher negative change. As indicated in the table, society B experienced a more substantial increase in the initial population, amounting to 624.44\% compared to society A. Similarly, the male population in society B exhibited a greater surge by 588.1\% in contrast to society A. Moreover, society B demonstrated a significant upsurge in the female population, showing a rise of 656.25\% in comparison to society A. The available volume of food (plants) in the Northeast region witnessed a marginally higher decline in society B by 0.17\% compared to society A. On the other hand, society A exhibited a greater reduction in the volume of available food (plants) in the Northwest region by 1.86\% compared to society B. Furthermore, the decrease in the volume of available food (plants) in the Southeast region was more pronounced in society A by 1.71\% than in society B. Finally, society A displayed a higher decrease in the volume of available food (plants) in the Southwest region by 2.17\% in contrast to society B.

\section{Conclusion}

The intricacies of our world unfold within the framework of complex adaptive systems, from the fundamental interactions of quarks to the boundless complexity of galaxies. Each level of organization, from subatomic particles to entire societies, exhibits a unique form of intelligence, resulting in a complex and interconnected tapestry of existence.Agent-based modeling has emerged as a powerful tool for comprehending the multifaceted nature of complex adaptive systems. Positioned within the expansive domain of artificial intelligence, this methodology enables researchers across disciplines to construct intelligent agents and simulate their interactions within a specific environment. The bottom-up simulation approach provides a robust framework for exploring and understanding the intricacies of these systems.

The study of society as a complex adaptive system exemplifies its dynamic nature, marked by a collective intelligence that transcends the individual elements. This exploration has revealed several crucial insights. Firstly, the capacity of complex adaptive systems to acquire and consolidate knowledge sets them apart from more straightforward systems. Secondly, their properties exhibit nonlinear changes over time, emphasizing the nontrivial nature of their evolution. Thirdly, the emergent behavior of these systems results from the intricate interplay of all constituent elements, rather than a mere sum of their individual contributions. Lastly, even slight alterations in a society's past can yield profound implications for its future, showcasing the significance of historical context and memory. Consequently, any analysis or comparison of societies must be contextualized within the historical narrative and collective memory of their inhabitants. Neglecting this critical aspect could lead to an incomplete understanding of their present status and trajectory. It is imperative for policymakers to consider the historical fabric of the society in question and develop policies that align with this understanding. In this pursuit, agent-based modeling stands as an invaluable tool, offering the potential for informed policy formulation and implementation.

\bibliographystyle{plain}
\bibliography{Sources}

\end{document}